\documentclass[aps,prl,showpacs,twocolumn,superscriptaddress,floatfix]{revtex4}

\usepackage{epsfig}
\usepackage{amsmath,amssymb,wasysym,epsfig,capt-of,ifthen,calc}
\usepackage{bbm}
\usepackage{latexsym}   
\usepackage{setspace}   
\usepackage{array}      
\usepackage{delarray}   
\usepackage{afterpage}
\usepackage{graphicx}
\usepackage{dcolumn}
\usepackage{bm}
\usepackage{array}
\usepackage{hyperref}
\usepackage{float}
\usepackage{supertabular}
\usepackage{longtable}
\usepackage{subfigure}
\newcommand{\be}{\begin{equation}}
\newcommand{\ee}{\end{equation}}
\newcommand{\bea}{\begin{eqnarray}}
\newcommand{\eea}{\end{eqnarray}}

\bibstyle{apsrev}

\begin{document}

\title{Attractor and Basin Entropies of Random Boolean Networks Under Asynchronous Stochastic Update}

\author{Amer Shreim}
\affiliation{Complexity Science Group, Department of Physics and Astronomy,
University of Calgary, Calgary, Alberta, Canada, T2N 1N4}

\author{Andrew Berdahl}
\affiliation{Complexity Science Group, Department of Physics and Astronomy,
University of Calgary, Calgary, Alberta, Canada, T2N 1N4}
\affiliation{Department of Ecology \& Evolutionary Biology, Princeton University, Princeton, NJ, USA, 08544}

\author{Florian Greil}
\affiliation{Arbeitsgruppe Komplexe Systeme, Institut f\"ur
  Festk\"orperphysik, Technische Universit\"at Darmstadt, D-64289 Darmstadt,
  Germany}
\altaffiliation{Current address: Climate Sciences, Alfred-Wegener-Insitute, Germany, D-27570 Bremerhaven}

\author{J\"orn Davidsen}
\affiliation{Complexity Science Group, Department of Physics and Astronomy,
University of Calgary, Calgary, Alberta, Canada, T2N 1N4}

\author{Maya Paczuski}
\affiliation{Complexity Science Group, Department of Physics and Astronomy,
University of Calgary, Calgary, Alberta, Canada, T2N 1N4}

\date{\today}

\begin{abstract}
We introduce a numerical method to study random Boolean networks with
  asynchronous stochastic update.  Each node in the network of states
  starts with equal occupation probability and this probability distribution 
  then evolves to a steady state.
  Nodes left with finite occupation probability determine the attractors and the
  sizes of their basins.  As for synchronous update, the basin entropy
  grows with system size only for critical networks, where the
  distribution of attractor lengths is a power law.  We determine
  analytically the distribution for the number of attractors and basin
  sizes for frozen networks with connectivity $K=1$.
\end{abstract}

\pacs{05.45.-a, 89.75.-k, 89.75.Fb, 89.75.Da}
\maketitle

Networks have proven to be useful in many fields. In particular for biological high-throughput experiments to construct regulatory networks there are immense challenges how to interpret their behavior.
Relevant criteria
for models include conceptual simplicity, computational tractability,
and robustness to uncertainties in the data. Boolean networks are
candidates for representing classes of behaviors observed
in large regulatory networks (see e.g. Ref.~\cite{cosentino2009}).
Biochemical details like reactions rates and concentrations (which are
often unknown) are discarded for a simpler description in which genes
or functional sets of genes are ``on'' or ``off''; for reviews see~\cite{aldanaRBN2003,bornholdt2008boolean}.  
 In addition to describing specific small regulatory
networks~\cite{ AlbertCellScience,li2004yeast}, Boolean networks have
been used to represent other complex dynamical systems such as
neural~\cite{rosen-zvi}, or evolving~\cite{paczuski-bassler2000} networks.

As canonical examples of disordered systems, random Boolean networks (RBNs)~\cite{Kauffman1969} have attracted the attention
of physicists over four
decades~\cite{aldanaRBN2003,drosselRBNreview}.  RBNs are directed graphs
consisting of $N$ Boolean elements, where each element receives input
from $K$ distinct elements. The value of the $i^{th}$ element,
$\sigma_i$, evolves according to a random Boolean function of its $K$
inputs: $ \sigma_i(t+1) =
f_i(\sigma_{i_1}(t),\sigma_{i_2}(t),\ldots,\sigma_{i_K}(t)),
$ where $\sigma_{i_j}(t)$ is the value of the $j^{th}$ input to
element $i$ at time $t$.  Here we choose the function $f_i$
to be zero or one with equal probability. The functions $f_i$
are fixed for each realization of the RBN.

In the classical RBN (CRBN), elements evolve simultaneously according
to a globally synchronized clock. States separate into transient
states, which cannot be reached more than once under the dynamics, and
attractor states, which can be reached infinitely often in the long time
limit.  Attractor states form closed loops in state space.
The state space as a whole divides into non-overlapping
partitions, or basins of attraction.  Transient states within a given
basin all reach the same attractor at long times.  By considering
small perturbations to states within an annealed approximation, Derrida and Pomeau~\cite{DerridaPomeau} found that $K=2$ CRBNs are
critical: they separate ordered $(K<2$) and chaotic $K>2$
phases, where distances between nearby
trajectories vanish or diverge, respectively.  Numerous investigations have focused on the lengths and numbers
of attractors and their basins of attraction~\cite{aldanaRBN2003,drosselRBNreview, Drossel2005PRE,
  Drossel2005PRL, samuelsson2003RBN, bastollaparisi, berdahl09,
  bhattaPRL}.  In particular, Krawitz and Shmulevich~\cite{ShmulevichEntropyRBN} found that the basin entropy, which
measures the variations in the size of the partitions of state space, increases with system size only for critical CRBNs.

However, most real world systems do not evolve according to a globally
synchronized clock, leaving serious doubts on the applicability of
CRBNs~\cite{harvey1997time, Bornholdt2005stable,glass98} as useful models.  Here we consider the opposite extreme and study asynchronous, stochastic random Boolean networks (ARBNs), where at each time step, a single randomly selected element is updated.  Previous studies have found fundamental differences between CRBN and ARBN ensembles. For instance
Ref.~\cite{greil2005dynamics} argued, using analytic methods, that shifting from synchronous to asynchronous dynamics drastically reduces the number of attractors for connectivity $K=2$.

Here we show, using a novel computational method, that certain
features of RBNs are the same for both extremes and
thus are generic with respect to the dynamics.  In particular the
basin entropy increases with system size only for critical
ARBNs. Hence it is a robust detector of critical
behavior for both CRBNs and ARBNs.  Further, the distribution of
attractor lengths is a power law for $K=2$ networks in both cases.
Finally, we show that all attractors in a $K=1$ ARBN have the
same length, and  all their basins have the same size.
We use this result to derive analytically the distribution of the
number of attractors and the average basin entropy for ensembles of 
$K=1$ ARBNs as a function of system size $N$.

An attractor in a discrete system with non-deterministic dynamics, such as an
ARBN, is also a subset of all possible states that can be reached
infinitely often in the long time limit starting from a random initial
condition. Mathematically, an attractor is a set of states such that starting from 
any state within it: (a)  all other states of the
attractor can be reached and (b) these are the only states that can be reached.
States that do not belong to an attractor are transient.
Since transient states may form loops in the state space network (SSN), our definition is
not the same as in Ref.~\cite{greil2005dynamics}.  We define the
normalized size of an attractor's basin to be the chance of reaching
that attractor starting from a randomly chosen initial
state. Note that unlike CRBNs, an ARBN can reach different attractors starting from the same initial state, i.e. the
transient states in different basins can overlap.  In addition,
attractors do not necessarily form simple loops. However, different
attractors are still composed of non-overlapping subsets of states.

We find attractors and their basins using the following method: First we
construct the entire SSN of ${\cal N}=2^N$ states for an
ARBN of size $N$, by representing the states as
nodes. A directed link, with weight $1/N$, points from a node to each
of its $N$ images reached by choosing, in turn, one element for
update.  An image can be the state itself or some other state with
Hamming distance one away. If some among these $N$ images are
identical, we add the weights of the corresponding links.  Second, our
algorithm initially assigns the same occupation probability  $\rho=1/{\cal N}$ to each
node.  
Then the algorithm updates this probability distribution on the SSN in parallel by dividing and moving the
entire occupation probability on each node along its outgoing links according to the
weights.  The algorithm repeats this step until the probability distribution is stationary~\footnote{In practice, our criteria for stationarity requires
  that for each node in the SSN, the change in its occupation probability after an update is
  smaller than $\epsilon$. Here we use $\epsilon=10^{-6}/{\cal N}$.}.
By definition, the occupation probability $\rho$ vanishes for transient states, but remains finite for states on attractors.  We identify different attractors as subsets of these latter states that are dynamically connected~\footnote{To check for consistency, we tested whether there were states identified erroneously  as
 transient that could be reached from an attractor. This was never observed.}.  The value of $\rho$ for each state is the chance
of observing that state in the long time limit after starting the ARBN
 from a random initial state. The sum of the densities for all
states on an attractor is its normalized basin size.

Before presenting our findings using this method, we first discuss the
case $K=1$, where we derive analytical results and compare them with
results from the SSN method, finding complete agreement.  The
structure of a $K=1$ RBN is particularly simple: Boolean elements form loops
with trees rooted in them.  Only four Boolean functions exist to update
elements: copy, invert, force to zero and force to one.  Relevant
components in CRBNs determine the number and lengths of attractors as
well as their basin sizes~\cite{flyvbjerg1988esk,drosselRBNreview}.
This is also true for ARBNs.  Loops of elements with at least one
forcing function become frozen eventually, as do elements on trees
rooted in frozen loops. Hence, relevant components are loops that
contain no forcing functions.  They are even or odd.  Even (odd) loops
have an even (odd) number of invert functions. Their number in ARBN
$i$ is $n_{even}(i)$ ($n_{odd}(i))$.

Without loss of generality, consider that all functions on even loops
are copy functions. Then the two stable states are all zeroes
$(00...00)$ and all ones $(11...11)$.  Hence a single even loop
generates two point attractors. By symmetry, the two point attractors
split the SSN into two equal parts. Hence, for an ARBN with
$n_{even}$ loops and no odd loops, the number of attractors is $2^{n_{even}}$. Each
has a basin of size $2^{N-n_{even}}$.

Odd loops are equivalent to a loop with one invert function, the rest
being copy.  Assume that the loop is in state ${\mathbf S_0} =
\{0\ldots0\}$ and call the element with the invert function
$\eta$.  ${\mathbf S_0}$ changes only when $\eta$ is chosen for update.  Then,
the new state ${\mathbf S_1}$ does not change until the neighbor on the loop
downstream from $\eta$ is chosen. Following this argument, we conclude
that a single odd loop of length $L$ in an ARBN has 
 $2 L$ different states it can be in.  An ARBN with $n_{odd}$ odd loops of lengths
$L_1, L_2, \ldots, L_{n_{odd}}$, each loop $j$ having $s_j$ unfrozen elements downstream of it, has one attractor with $\prod_{j= 1}^{n_{odd}}  2 L_{j} 2^{s_{j}}$ states.  Thus for ARBN $i$ that contains
$n_{even}(i)$ and $n_{odd}(i)$ loops, the number of attractors is
$A_i=2^{n_{even}(i)}$.  Since the attractor from the odd loops enters equally into all basins, each attractor $\alpha$ drains a basin of size $b_{\alpha} =2^{N-n_{even}(i)}$.
Indeed, we observe numerically that in a given single ARBN the number of states in each attractor is the same for all
attractors and that all basin sizes are also the same.

\begin{figure}[!h]
\begin{center}
\includegraphics*[width=\columnwidth]{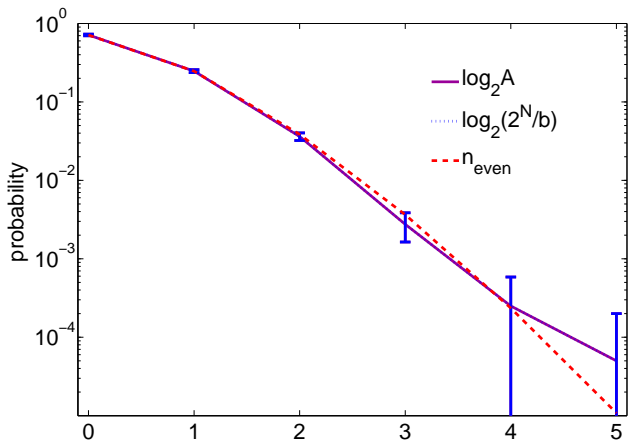}
\caption{\label{K1Analytical} The distribution of the number of even
  loops $P(n_{even})$, obtained using $Q({\mathbf m})$ -- compared to the
  distribution of the number of attractors, $P(\log_2 A)$, and the
  distribution of inverse basin sizes, $P(\log_2 \frac{{\mathcal
      N}}{b})$, obtained using our SSN method for $K=1$ ARBNs with
  $N=16$.  The number of realizations is $2\times10^4$ and the error bars
correspond to two standard deviations.}
\end{center}
\end{figure}

We now use results of Ref.~\cite{flyvbjerg1988esk} to find the
distribution of the number of even loops, $P(n_{even})$, over ensembles of ARBNs.  These
authors derive the probability, $Q({\mathbf m})$, to observe the vector
${\mathbf m} = \{m_1, m_2, \ldots\}$ relevant loops of lengths $L =
\{1,2, \ldots\}$ in a $K=1$ RBN. In our case this probability reads
\begin{equation}
Q({\mathbf m}) = \left(\frac{1}{2}\right)^{\hat{m}} \frac{N!} {(N-\hat{m})! N^{\hat{m}} }
\frac{1}{2}(1+\hat{m}/N) \prod_{L=1}^{\infty} \frac{L^{-{m_L}}}{m_L!},
\nonumber
\label{ProbRelLoops}
\end{equation}
where $\hat{m} = \sum_{L=1}^{\infty} m_L L$.
The probability $P(n)$ to have  $n$ relevant
loops (both even and odd) is $ P(n) = Q({\mathbf m} | \sum_L m_L = n)$.
$P(n_{even})$ is obtained from $P(n)$ as
\begin{equation}
P(n_{even}) = \sum_{n = n_{even}}^{\infty} P(n) {n \choose n_{even}} \left(\frac{1}{2}\right)^n \quad .
\nonumber
\end{equation}
These expressions do not have closed form solutions, but can be
evaluated numerically.  Fig.~\ref{K1Analytical}
shows $P(n_{even})$ evaluated from $Q({\mathbf m})$.  Noting that $P(n_{even}) =
P(\log_2 A) = P(\log_2 \frac{{\mathcal N}}{b})$,
Fig.~\ref{K1Analytical} also shows $P(\log_2 A)$ and $P(\log_2 \frac{{\mathcal N}}{b})$,
obtained using the numerical SSN method described next, finding  agreement.

We now consider $K\geq1$, and distinguish two measures for the length of
attractors.  The first measure, $l_{\alpha}$, is the actual number of 
states on attractor $\alpha$.  The second measure, $\lambda_{\alpha}$, takes
  into account how often states  are visited.
Defining the stationary occupation probability of state
  $j$ on attractor $\alpha$ as $\rho_{j,\alpha}$, the conditional
  probability of that state given that the dynamics reaches attractor $\alpha$ is
\begin{equation}
\varrho_{\alpha j} = \frac{\rho_{\alpha j}}{\sum_{j=1}^{l_\alpha} \rho_{\alpha j}}\quad .
\nonumber
\end{equation}
The attractor entropy, $s_{\alpha}$, determines  $\lambda_{\alpha}$ through:
\begin{equation}
\ln \lambda_{\alpha}=
s_{\alpha} = - \sum_{j=1}^{l_\alpha} \varrho_{\alpha j} \log \varrho_{\alpha j} \quad .
\nonumber
\end{equation}
If all states on attractor $\alpha$ are visited equally often then
$\lambda_{\alpha}=l_{\alpha}$.  If, on the other hand, one state dominates,
$\lambda_{\alpha}\rightarrow 1$.  We observe a strong correlation between $l_{\alpha}$
and $\lambda_{\alpha}$ (data not shown). Since we find clearer -- although not
qualitatively different --  behavior using $\lambda$, we present those
results below.

\begin{figure}[!h]
\begin{center}
\subfigure[]{\includegraphics*[width=\columnwidth]{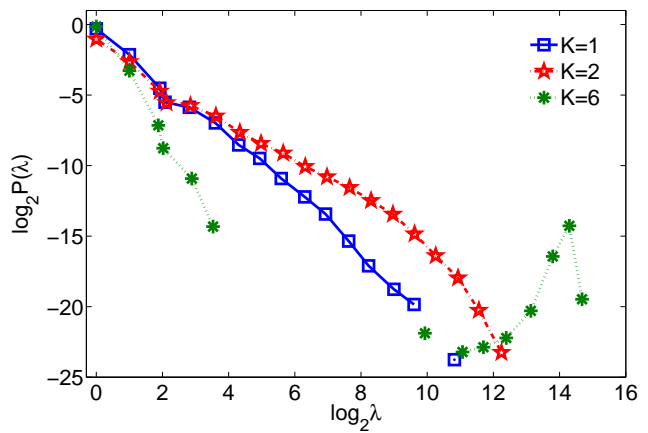}}\\
\subfigure[]{\includegraphics*[width=\columnwidth]{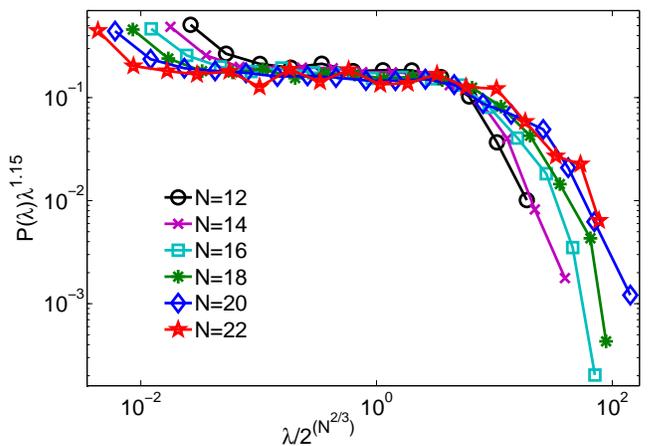}}
\end{center}
\caption{\label{AttractorAndBasinStatistics} (a) $P(\lambda)$ for $K=1,2,6$
  and $N=16$.  Both $K=1$ and $K=2$ ARBNs show a broad distribution while all
  attractor lengths for $K=6$ are either of order unity or of order
  $2^N$.
  (b) Rescaled $P(\lambda)$ {\it vs.} rescaled $\lambda$ for $K=2$
  and $N = 12 - 22$. The approximate collapse is consistent with a power law decay up to
  a cutoff which is a stretched exponential in $N$.}
\end{figure}

Fig.~\ref{AttractorAndBasinStatistics}a shows the probability density function
for the length of attractors, $P(\lambda)$, for $N=16$.  It is broad
for $K=1$ and $K=2$.  For $K=6$, $\lambda$ is either
of order unity or of the SSN size $\mathcal N$. This is true for all
$K>2$ although subleading effects dominate for $2<K<6$, and
no clear separation appears for this ${\mathcal N}$.
A rescaled $P(\lambda)$ is shown in Fig.~\ref{AttractorAndBasinStatistics}b. The cutoff in the length $\lambda_c \sim 2^{N^{2/3}}$
is suggested by the dimensional arguments in Ref.~\cite{greil2005dynamics} and the data is consistent with a power law decay
$P(\lambda)\sim \lambda^{-\tau_\lambda}$ with $\tau_\lambda = 1.15 \pm
0.05$. However, the cutoff changes shape for different $N$ over the considered
range of system sizes so the system is not yet in a regime where subleading corrections
can be ignored.

\begin{figure}[!h]
\includegraphics*[width=\columnwidth]{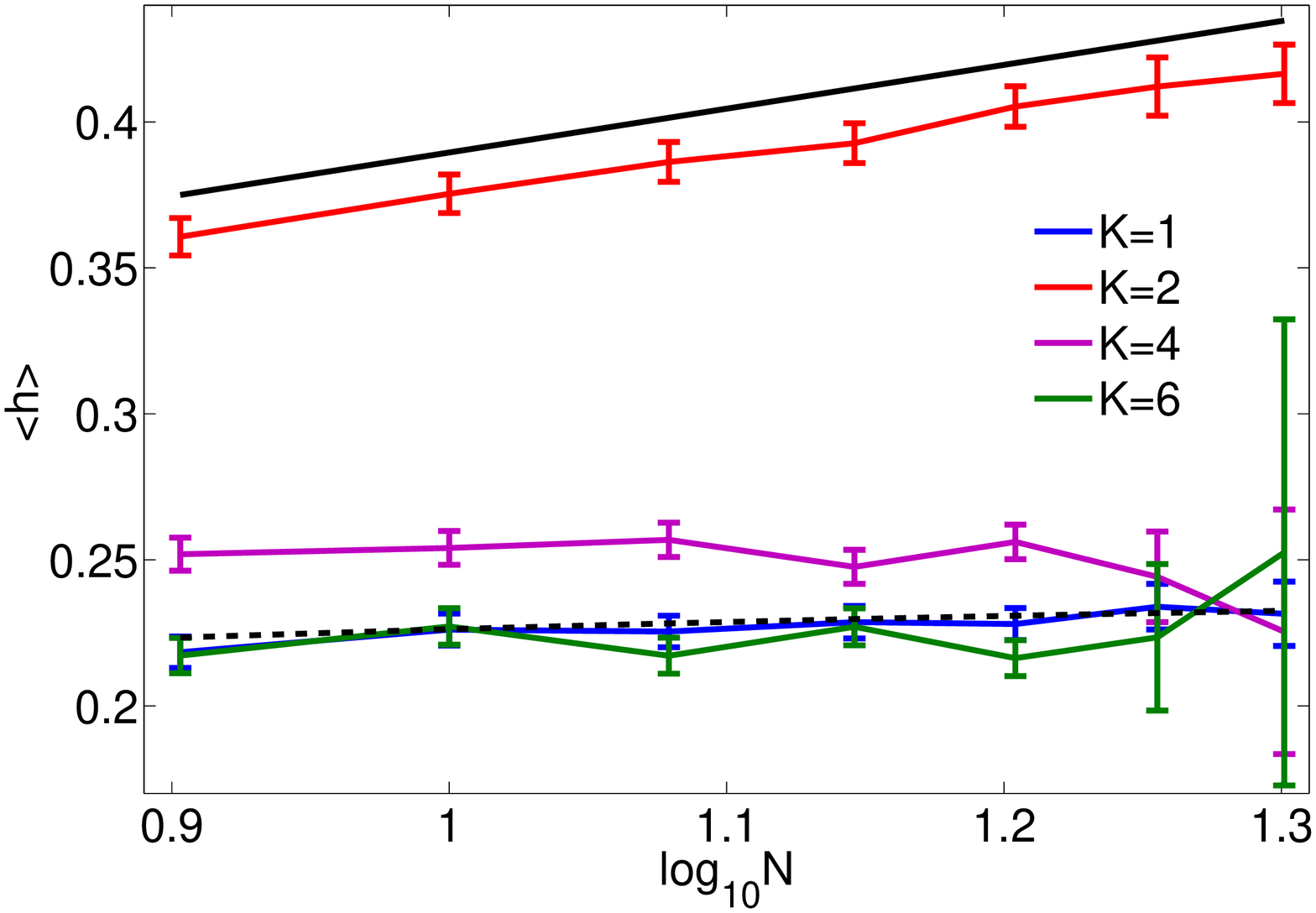}
\caption{\label{BasinEntropy} The  basin entropy $\langle h \rangle$
  for ARBNs with various $K$, and $N=8 -20$.
  $\langle h \rangle$ grows with $N$ only for $K=2$.  The dashed  line shows  $\langle h \rangle$ for $K=1$ derived from $Q({\mathbf m})$. The solid black line has a slope of $0.15$ and is a guide for the eye.}
\end{figure}

The probability, $p_{\alpha}$, to reach attractor $\alpha$ from a
random initial state is the sum of all occupational probabilities,
$p_{\alpha}=\sum_{j=1}^{l_{\alpha}} \rho_{\alpha j}$ over  states on
$\alpha$.  Its basin size, $b_{\alpha}={\mathcal N}p_{\alpha}$.
Ref.~\cite{ShmulevichEntropyRBN} proposed that the average basin entropy $\langle h\rangle$, which gives a summary measure
for the variation in basin sizes, is a measure of complexity and criticality in
CRBNs.  The basin entropy of RBN $i$ is
\begin{equation}
 h_i = - \sum_{\alpha =1}^{A_i} p_{\alpha} \ln
p_{\alpha} \quad ,
\end{equation}
and the average basin entropy over the ensemble is $\langle h \rangle=
\frac{1}{R}\sum_{i}^R h_i$ where $R$ is the number of realizations.
Fig.~\ref{BasinEntropy} shows the average basin entropy $\langle h
\rangle$ for ARBNs with $K = 1,2,3,6$ and $N = 8$ to $20$.  This figure
shows that $\langle h \rangle$ grows only for $K=2$ while remaining
constant for $K \neq 2$.  For $K=1$, Fig.~\ref{BasinEntropy} also shows
$\langle h\rangle$ derived from $Q({\mathbf m}$), which agrees perfectly with the SSN method. For CRBNs~\cite{ShmulevichEntropyRBN}, as
well, the basin entropy was shown to grow with system size only for the critical
case $K=2$ -- over the same range of system sizes.

Ref.~\cite{berdahl09} showed that the distribution of attractor
lengths in CRBNs depends on the sampling scheme.  For CRBNs with $K>2$ the
distribution obtained by counting all attractors is a power law,
while that obtained by randomly sampling initial states is not.  For
$K=2$ both distributions are power laws.  This is due to a lack of
correlation between the size of an attractor and the size of its basin
for $K=2$, while for $K>2$  the basin size grows linearly with
attractor length.  In the ARBNs studied here, the basin size varies
little with attractor length for all $K$, so the two distributions are
indistinguishable over the range of system sizes $8\leq N\leq 20$.

As in Ref.~\cite{ShmulevichEntropyRBN}, a method based on exact
enumeration of state space creates a severe restriction on the system sizes that can be studied.  Indeed for $K=2$ CRBNs, the mean number of attractors grows
faster than any power law with system size~\cite{samuelsson2003RBN},
but one cannot reach this asymptotic result for CRBNs in the range of
$8\leq N\leq 20$.  The growth we observed for CRBNs is much slower (data not
shown).  For ARBNs, the mean number of attractors increases as a power law with
system size for $K=2$~\cite{greil2005dynamics}.  We also do not observe this asymptotic
result for the range of system sizes studied (data not shown).  Hence, it is not possible to extract an asymptotic growth
law for the size dependence of the basin entropy for critical networks
using this method in either case.

To summarize, we have developed a method to find the long-time dynamics of ARBNs. It relies on the flow of the occupation probability over the complete network of possible states until stationarity is reach. We focussed on how the
space of  states divides into overlapping basins of
attraction and measured the fluctuations of these fuzzy partitions in
ensembles of ARBNs.  Despite the fundamental difference between ARBNs
and classical RBNs, we find that in both cases the basin entropy increases with
system size only for critical networks.  This suggests
that ``superuniversal'' features exist for the fluctuations in the
structure of state space of RBNs -- these are invariant with respect to the specific
dynamics.  As the dynamics for real world systems most likely lies in
between these two extremes, our work suggests that the basin entropy may
be a relevant and robust signifier of criticality, irrespective of the
specific dynamics.  Finally, we showed that all attractors in an ordered
$K=1$ ARBN have the same length, and that all their basins have the same
size.  We used this to derive analytically the distributions of both basin entropy and the
number of attractors in $K=1$ ARBNs as a function of system size.

\end{document}